\documentclass[lettersize,journal]{IEEEtran}
\usepackage{amsmath,amsfonts}
\usepackage{algorithmic}
\usepackage{algorithm}
\usepackage{cite}
\usepackage{array}
\usepackage{color}
\usepackage[caption=false,font=normalsize,labelfont=sf,textfont=sf]{subfig}
\usepackage{textcomp}
\usepackage{stfloats}
\usepackage{url}
\usepackage{verbatim}
\usepackage{booktabs}
\usepackage{graphicx}
\usepackage{cite}
\usepackage{caption}
\begin{document}

\title{SemProtector: A Unified Framework for Semantic Protection in Deep Learning-based Semantic Communication Systems}

\author{Xinghan Liu, Guoshun Nan, Qimei Cui, Zeju Li, Peiyuan Liu, Zebin Xing, Hanqing Mu, \\ Xiaofeng Tao, Tony Q.S. Quek,~\IEEEmembership{~Fellow,~IEEE}

\thanks{
This work was supported by National Key R\&D Program of China (Grant No.2022YFB2902200).
Xinghan Liu, Guoshun Nan, Zeju Li, Qimei Cui, Peiyuan Liu, Zebin Xing, Hanqing Mu and Xiaofeng Tao are with National Engineering Research Center for Mobile Network Technologies, Beijing University of Posts and Telecommunications, Beijing 100876, China. (\textit{Corresponding author: Guoshun Nan.})}
\thanks{T. Q. S. Quek is with the Singapore University of Technology and Design, Singapore 487372, and also with the Yonsei Frontier Lab, Yonsei University, South Korea.}}

\markboth{Journal of \LaTeX\ Class Files,~Vol.~14, No.~8, August~2021}%
{Shell \MakeLowercase{\textit{et al.}}: A Sample Article Using IEEEtran.cls for IEEE Journals}


\maketitle

\begin{abstract}
Recently proliferated semantic communications (SC) aim at effectively transmitting the semantics conveyed by the source and accurately interpreting the meaning at the destination. While such a paradigm holds the promise of making wireless communications more intelligent, it also suffers from severe semantic security issues, such as eavesdropping, privacy leaking, and spoofing, due to the open nature of wireless channels and the fragility of neural modules. Previous works focus more on the robustness of SC via offline adversarial training of the whole system, while online semantic protection, a more practical setting in the real world, is still largely under-explored. To this end, we present SemProtector, a unified framework that aims to secure an online SC system with three hot-pluggable semantic protection modules. Specifically, these protection modules are able to encrypt semantics to be transmitted by an encryption method, mitigate privacy risks from wireless channels by a perturbation mechanism, and calibrate distorted semantics at the destination by a semantic signature generation method. Our framework enables an existing online SC system to dynamically assemble the above three pluggable modules to meet customized semantic protection requirements, facilitating the practical deployment in real-world SC systems. Experiments on two public datasets show the effectiveness of our proposed SemProtector, offering some insights of how we reach the goal of secrecy, privacy and integrity of an SC system. Finally, we discuss some future directions for the semantic protection.
\end{abstract}


\section{Introduction}
Existing wireless communication systems, which are built on the classical Shannon information theory, primarily focused on how to accurately and effectively transmit raw bits over channels from a source to a destination. In the past decades, the achieved transmission rate has been improved tens of thousands of times to meet the increasing traffic demands, gradually driving a wireless system capacity to the Shannon limit. Meanwhile, various newly emerged mobile applications, such as virtual reality (VR) and Human-to-Machine (H2M) communications, tend to require more intelligence among different parties, as well as interpretations of the received information to make smart decisions. This motivates us to rethink a more intelligent and a more efficient wireless communication paradigm beyond the Shannon approach. 

Recently proliferated semantic communications (SC) systems \cite{xie2020deep,han2022semantic,nan2023udsem,xiao2022imitation} have shown great potential in transmitting the semantics conveyed by the source and accurately interpreting the meaning at the destination. Such a novel communication paradigm greatly facilitates the aforementioned mobile applications such as VR and H2M. Specifically, the transmitter of an SC system relies on neural networks to extract semantic information from an input and then sends compressed semantic symbols to noised wireless channels. At the receiver side, the system reconstructs the received semantics and interprets the semantics to make intelligent decisions.  

Although promising, SC systems may suffer from more challenging security issues compared to traditional wireless communications, such as eavesdropping, tampering, and spoofing due to the broadcast nature of wireless communications and the fragility of deep neural networks\cite{nan2023physical,qin2023securing}. First, anyone within the physical communication range of a transmitter can receive the wireless signal and potentially decode the symbols. As an SC system primarily focuses on the transmission of the meaning conveyed at source rather than accurate raw bits, it facilitates eavesdroppers to carry out malicious attempts to semantics, such as inference attacks and adversarial attacks\cite{sadeghi2019physical}. Second, as an SC system is learned from a large volume of data examples, which may consist of sensitive information from  the personal knowledge base, such as the user's health status and service access history, the system may unintentionally reveal sensitive information in semantic representations to be sent to wireless channels, potentially leading to privacy leaking of the input semantics\cite{shokri2017membership}. Third, an SC system is vulnerable to adversaries due to the blind spots of neural models. The adversarial perturbations are able to mislead the system to make incorrect semantic interpretations.

Existing works focus more on the robustness of SC via offline adversarial training of the whole system, mitigating the blind spots of the neural models with adversary examples\cite{hu2022robust,peng2022robust}. These studies are able to harden the system against adversarial attacks. However, they may not be practically deployed in real-world scenarios as it is infeasible to improve the robustness of neural models by interrupting a communication system for re-training. Furthermore, these studies mainly aim to defend against a single attack and hence are unable to protect an SC system against various threats. Therefore, it is still unclear how an online SC system can be simultaneously protected against various malicious attempts such as eavesdropping, tampering and spoofing.

\begin{figure*}[ht]
	\centering
	\includegraphics[scale=0.48]{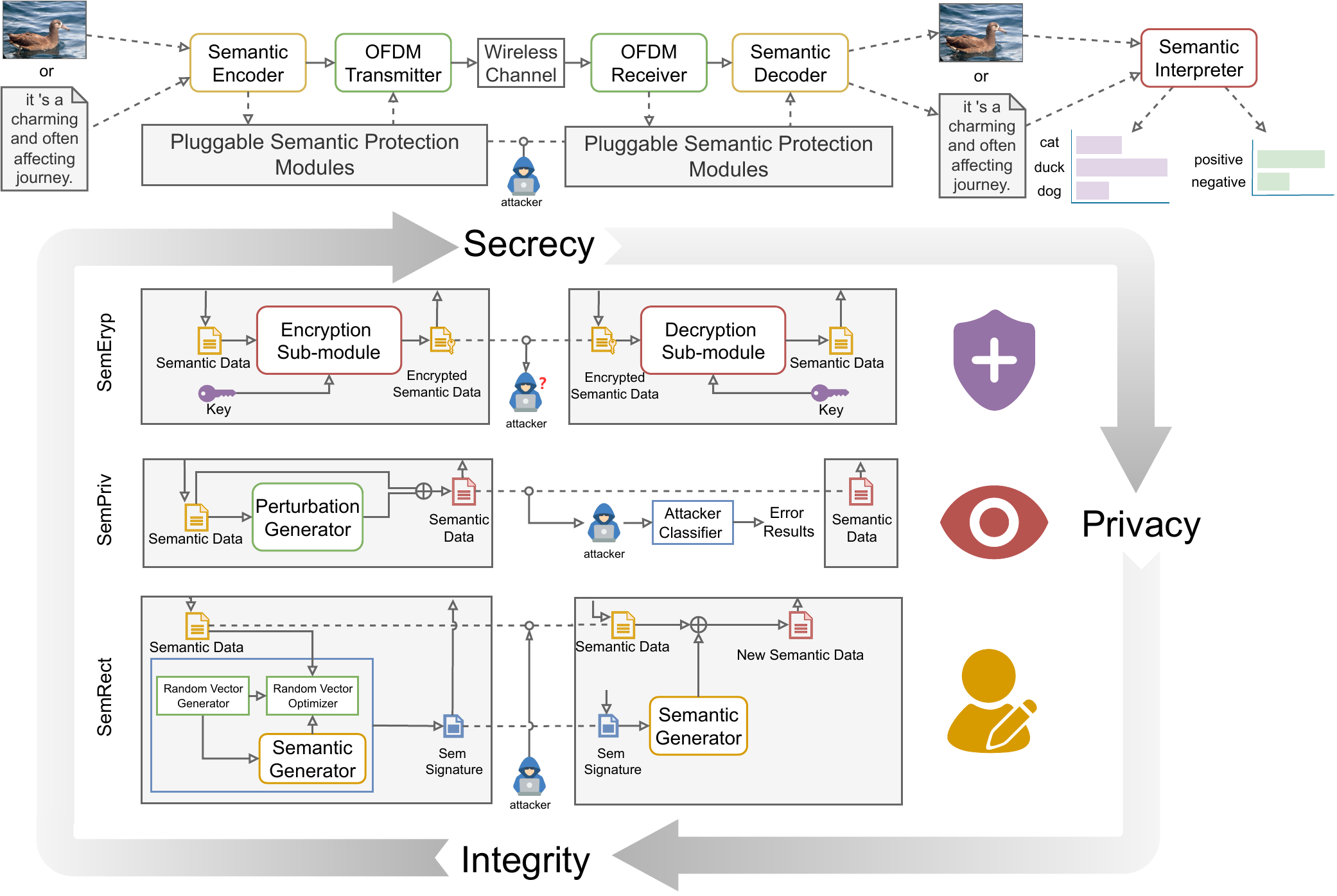}
	\caption{Overview of our proposed SemProtector framework. The top of the figure demonstrates an end-to-end semantic communication system, equipped with our hot-pluggable semantic protection modules residing in the transmitter and receiver. At the bottom of the figure are hot-pluggable modules, including SemEryp, SemPriv, and SemRect, with the goal of secrecy, privacy, and integrity, respectively. The three modules can be flexibly assembled based on the customized protection requirements.}
	\label{fig1:fig1}
\end{figure*}

To this end, we present SemProtector, a unified framework that aims to secure the semantic transmission and interpretation of SC systems with three hot-pluggable semantic protection modules, including SemEryp, SemPriv and SemRect. Specifically, SemEryp is able to encrypt semantics to be transmitted by an encryption at the transmitter, securing the semantics over an open wireless channel, and SemPriv aims to mitigate privacy leaking via a perturbation generator that can craft adversaries to distort the malicious actors. SemRect is capable of ensuring the semantic interpretation by generating a signature at the transmitter and then calibrating the distorted semantics via the signature at the receiver. We conduct experiments on two benchmarks.

In conclusion, our main contributions can be summarized in three-fold. First, we propose a unified framework to secure the online deep learning-based semantic communication systems with the goal of secrecy, privacy and integrity. Second, we present three hot-pluggable semantic protection modules under the unified framework. Our framework enables an existing online SC system to dynamically assemble the three modules to meet customized semantic protection requirements, facilitating the practical deployment in real-world SC systems. Third, we show the superiority of our method in securing semantic communications against various threats, giving some helpful insights for online semantic protections. We also discuss some future directions in the field.

\section{End-to-end Semantic Communication System}
\subsection{Overview}
In this section, we describe the end-to-end semantic communication system used in our framework and outline each component. We refer to the previous work JSCC-OFDM\cite{yang2022ofdm} and DeepSC\cite{xie2020deep} to implement our SC system. Our system involves a semantic encoder, an Orthogonal Frequency Division Multiplexing (OFDM) transmitter, an OFDM receiver, and a semantic decoder. We additionally introduce a semantic interpretation module based on Mobilenetv2\cite{sandler2018mobilenetv2}. We formulate our objective as the weighted sum of the data reconstruction loss and semantic interpretation loss, training the system under a multi-task learning paradigm. The top of Fig. \ref{fig1:fig1} demonstrates the proposed end-to-end SC system.

\subsection{Semantic Transmitter}
\noindent
{\textbf{Semantic Encoder: }} We feed the image or text to the semantic encoder to obtain the semantic representations. We refer to JSCC-OFDM/DeepSC to jointly consider the source encoding and channel encoding in our semantic encoder. 

\noindent
{\textbf{OFDM Transmitter: }} To make efficient use of the spectrum and reduce the computational overhead, we employ OFDM as our wireless transmission scheme, which is able to encode the semantic representations into multiple carrier frequencies without inter-symbol interference (ISI). It should be noted that the transmitter also supports the single-carrier OFDM mode with configurable parameters.

\subsection{Wireless Channel}
As discussed in end-to-end communications, deep neural networks are able to model the wireless physical channels, including additive white Gaussian noise (AWGN), the erasure channel, and the Rayleigh fading channel. In this article, we mainly consider the Rayleigh fading channel as it can better model the effect of a propagation environment for semantic communication systems.

\subsection{Semantic Receiver}

\noindent
{\textbf{OFDM Receiver: }}{The OFDM receiver takes received symbols as the input and performs inverse operations that have been done in the OFDM transmitter.}

\noindent
{\textbf{Semantic Decoder: }}{We feed semantic representations to the semantic decoder to reconstruct the input data, which jointly considers the channel and source decoding.}

\noindent
{\textbf{Semantic Interpreter: }}{This module takes the output of the semantic decoder as an input, and interprets semantics to make the corresponding decisions.}

\section{SemProtector Framework}

\subsection{Overview}
In this section, we introduce our proposed SemProtector framework, which is illustrated in Fig.\ref{fig1:fig1}. Our framework consists of three key modules including SemEryp, SemPriv, and SemRect. The goal of these three modules is to obtain the secrecy, privacy, and integrity of our SC system, respectively. We first outline the three modules as follows.

\subsubsection{SemEryp} The module aims to encrypt the semantic representations at the sender side and decrypt the information at the receiver side. Equipped with our SemEryp, an illegal user who can receive the wireless signal is unable to reconstruct the semantic data and hence confidentiality of semantics can be properly protected, while legitimate users can normally decode the signal.  

\subsubsection{SemPriv} This module takes the semantic representation as an input and then generates tiny perturbations that can mislead an eavesdropper to make an incorrect semantic interpretation. Meanwhile, such an operation will be transparent to legitimate users for interpretations. Hence the privacy leaking risks over the wireless channels can be properly mitigated.

\subsubsection{SemRect} The goal of this module is to ensure the accurate semantic interpretation of an SC system under the destructive physical layer adversarial attacks. It generates a high-dimension random vector based on semantic representation at the transmitter to calibrate the semantic data against adversarial attacks at the receiver. We term such a vector as a semantic signature. By doing so, we are able to protect the semantic integrity of the source to accurately understand the meaning for making decisions.

\begin{figure*}[!t]
	\centering
	\includegraphics[scale=0.37]{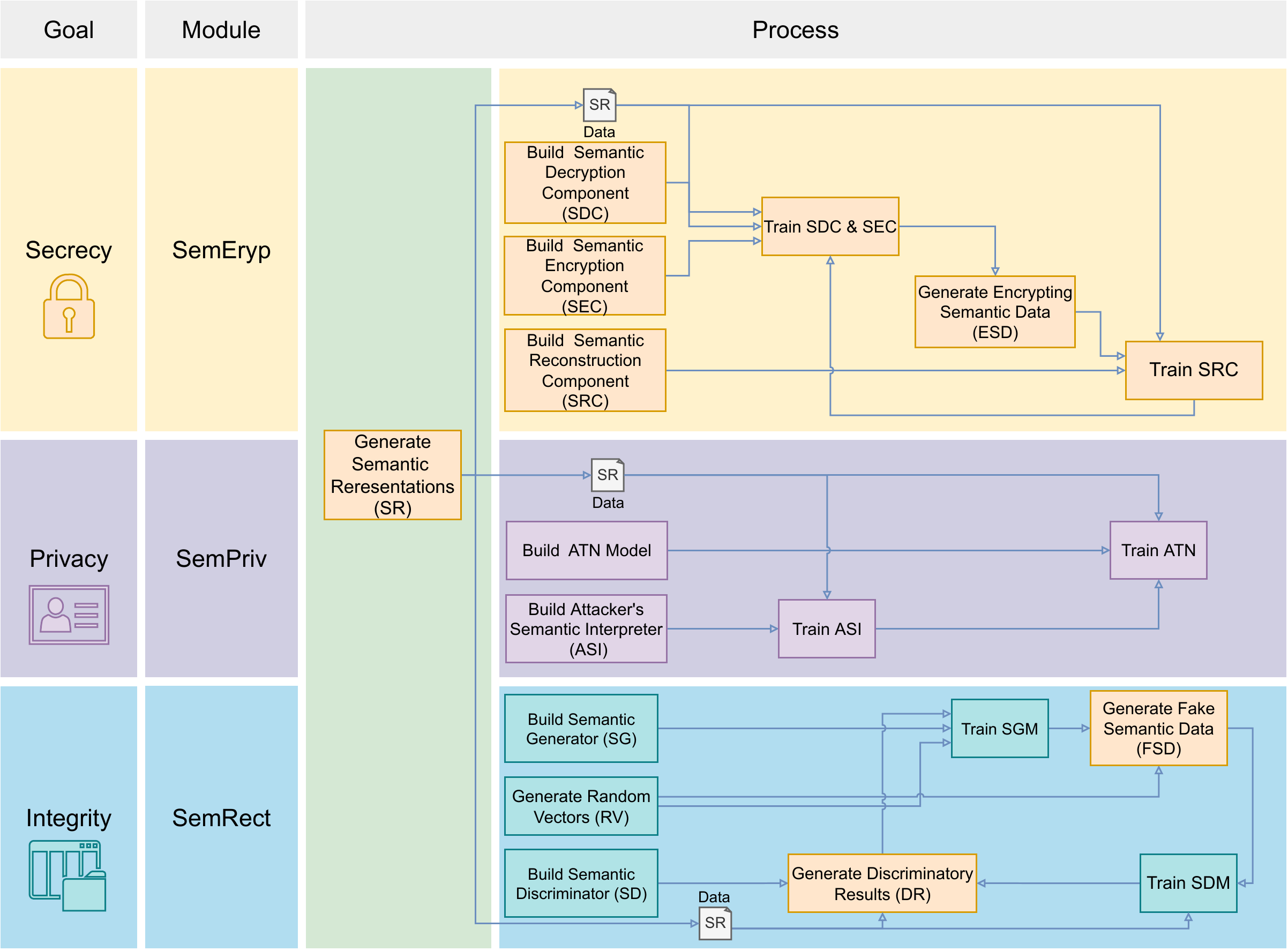}
	\caption{Training process of the proposed SemProtector, equipped with three semantic protection modules, including SemEryp, SemPriv, and SemRect. }
	\label{fig2:fig2}
\end{figure*}

We train these modules independently and then plug them into  the above end-to-end SC system for semantic protection. Our framework allows an existing online SC system to flexibly assemble the three modules to meet customized semantic protection requirements, greatly facilitating practical deployment in real-world SC systems. Fig.\ref{fig2:fig2} shows the training process of the three modules, and we detail them in the following parts.

\subsection{SemEryp}

The primary goal of SemEryp is to protect the confidentiality of the semantics underlying the input data. A malicious user can potentially reconstruct the input data based on semantics to carry out model inversion attacks \cite{fredrikson2015model,luo2023encrypted}, in case the eavesdropper acquires the knowledge of semantic decoder or trains a surrogate semantic decoder. This type of attack may lead to privacy leaking issues. We present a semantic encryption module to protect semantics over the air in our SemEryp.

Our SemEryp is composed of two components, i.e., semantic encryption (SEC) and semantic decryption (SDC).
SEC and SDC share the same neural structure, and both of them use multiple convolutional layers for encryption and decryption. As shown in the top of Fig \ref{fig2:fig2} colored in yellow, we feed the semantic representation of an input data and a semantic key to output encrypted semantic data, where the key is generated by a negotiation between the transmitter and the receiver. SDC takes the encrypted data and the semantic key as inputs to obtain the original semantics.

We demonstrate how we build our SemEryp with an adversarial training method. Specifically, we first introduce the semantic reconstruction component (SRC), which simply consists of multiple convolutional layers. We train SRC to recover the encrypted semantic data without the semantic key. Such a module can be regarded as a malicious user. Then we train SEC and SDC to defend against the attacks generated by SRC. We repeat the above procedure to achieve better confidentiality protections.

\subsection{SemPriv}
The primary goal of SemPiv is to protect the semantic privacy of inputs over an end-to-end SC system. We adapt  Adversarial Transformation Networks (ATN) to our semantic communication system to defend against attribute inference attacks (AIA), a destructive privacy attack that is able to infer hidden information of the semantics. An eavesdropper within the physical communication range of the transmitter can receive the wireless signal and potentially decode the semantics, leading to the privacy leakage of legitimate users.

We demonstrate how we train our SemPriv as follows, as shown in the middle of Fig.\ref{fig2:fig2} colored in purple. Specifically, we introduce an attacker's semantic interpreter (ASI) module. The module is a multilayer perceptron (MLP) and can be regarded as an eavesdropper. Then we feed the semantic representations to ATN to learn to craft the adversarial privacy perturbations by maximizing the cross-entropy loss of the ASI, so that the perturbation can mislead the eavesdropper to make an incorrect prediction. However, such a privacy perturbation will not affect the semantic interpretation of legitimate users, and hence the privacy leaking issue can be properly mitigated.

\subsection{SemRect}
Our SemRect aims to protect the integrity of semantics for accurate semantic interpretation at the receiver side. Here, an attacker relies on FGSM\cite{goodfellow2014explaining} to generate an adversarial perturbation against the semantic information to mislead the interpretations, where the perturbation will be added to the semantic data. Meanwhile, we also produce a random vector (RV), which is termed as a semantic signature, to calibrate such a perturbation, so that the impact of the attacks can be properly mitigated for legitimate users.

The detailed training process of SemRect is described as follows. We introduce a semantic generator (SG), which is adapted from the previous Defense-GAN. First, we train generative adversarial networks (GAN) to produce semantic representations based on a random vector. Such a process is illustrated at the bottom of Fig.\ref{fig2:fig2} colored in blue. We deploy the generator RV to both the sender and receiver. At the sender, RV learns to generate and then optimize semantic signatures. At the receiver, RV learns to calibrate the semantics with the signature, subtracting the adversarial perturbations for accurate semantic interpretations. By doing so, our SemRect is able to protect the integrity of semantics against adversarial perturbations over the end-to-end SC system.

\subsection{Summary}
In the above SemEryp, SemPriv and SemRect, we only select some attacks to demonstrate how each single module defends against a certain type of threat. By dynamically assembling these hot-pluggable modules, our SemProtector framework is able to defend against various attacks and meet the customized protection requirements, such as secrecy, privacy and integrity. Our unified framework is flexible for a further extension and can be generalized to defend against other attacks, facilitating semantic protection in a real-world SC environment. Due to the space limitation, we will not give more discussions for such an extension. 

\section{Experiments}
We conducted experiments on three popular datasets, MNIST, CIFAR10 and SST-2. The first two datasets are used for image classification, and the last one is used for text classification. We trained the SC system and the pluggable semantic protection modules on these three datasets and performed attacks on the SC system before and after adding the pluggable semantic protection modules to evaluate the effectiveness of SemProtector against various attacks.
Although we employ the Rayleigh fading channel in our SC system, our SemProtector can also be applied in other wireless environments, and we will not give the details due to the space limitation. We then detail the attack methods and evaluation metrics, and give some insightful conclusions based on our observations.
\subsection{Attack Methods}
\par{To evaluate the effectiveness of our framework, we re-implement three attack methods and outline them as follows.}
\begin{itemize}
    \item \textbf{Model Inversion Attacks:} In this attack, the attacker needs to intercept the semantic data and convert it to the original information. There are two ways for the attacker to get the original information by semantic data, one is to train the surrogate model and the other is to steal the semantic decoder of the receiver. Here we re-implement the second one, as it is more destructive. 
    \item \textbf{Attribute Inference Attacks:} For this attack, the goal of the attacker is to infer privacy information from the semantic data. Specifically, the attacker trains a semantic classifier that is targeted to interpret semantic data. In our implementation, the classifier trained by the attacker is an MLP. We use the accuracy of the attacker's classification of semantic data to evaluate the protection effectiveness of the attack.
    \item \textbf{Adversarial Attacks:} The goal of an adversarial attacker is to mislead the receiver by adversarial perturbations. Here we integrate FGSM to our SemRect module. We train the FGSM with a substitute model to mimic the SC system. The FGSM attacker is able to generate perturbations under a black-box setting and hence can be directly used in an online deep learning-based semantic communication system. 
\end{itemize}
\subsection{Evaluation Metrics}
We use Peak Signal to Noise Ratio (PSNR) and Structural Similarity
Index Measure (SSIM) to measure the impact of the optional pluggable semantic protection modules on the SC system. 
\begin{itemize}
 \item 
$\bold{PSNR}$ is a measure of the quality of an image or video. It is typically used to assess the fidelity of a compressed image or video file. The higher the PSNR, the better the quality of the image or video. 
\item 
$\bold{SSIM}$ is used to show how closely two images resemble one another. In contrast to PSNR, SSIM is more in accordance with how the human eye naturally perceives an image.

\end{itemize}
\subsection{Implementation Details}

We implement our SemProtector based on Pytorch, one of the most popular machine-learning libraries. We adapt JSCC-OFDM\cite{yang2022ofdm} and DeepSC\cite{xie2020deep} to our end-to-end SC system. Considering the computing resources and energy efficiency of mobile devices, we use Mobilenetv2\cite{sandler2018mobilenetv2} and TextCNN as the semantic interpreter of the SC system. For SemEryp, we use a simple neural network with five convolutional layers, which can encrypt and decrypt the semantic data very well at the same time taking into account the efficiency. For SemPriv, we use the modified ATN, which aims to craft semantic perturbations based on semantic data with six linear layers. For SemRect, we use the modified Defense-GAN, which can produce semantic signatures based on semantic data. These three hot-pluggable  protection modules can be flexibly assembled based on customized protection requirements.
\begin{table}[htbp]
\caption{Parameters in experiments.\label{tab:param}}
\resizebox{\linewidth}{!}{
\begin{tabular}{@{}lll@{}}
\toprule
Category &
  Parameter &
  Value \\ \midrule
Input &
  \begin{tabular}[c]{@{}l@{}}Shape(MNIST, CIFAR10,SST-2)\\ Num of class type (MNIST, CIFAR10,SST-2)\\ Num of train/test(MNIST)\\ Num of train/test (CIFAR10)\\ Num of train/test (SST-2)\end{tabular} &
  \begin{tabular}[c]{@{}l@{}}(1,28,27),(1,32,32),(1,64)\\ 10,10,2\\ 60,000/10,000\\ 50,000/10,000\\ 67.350/1,821\end{tabular} \\ \hline
ESC system &
  \begin{tabular}[c]{@{}l@{}}Num of residual block(image)\\ Num of transformer block(text)\end{tabular} &
  \begin{tabular}[c]{@{}l@{}}4\\ 2\end{tabular} \\ \hline
OFDM Module &
  \begin{tabular}[c]{@{}l@{}}Num of pilot symbols\\ Num of subcarriers per symbol\\ Length of Cyclic Prefix\\ Exponential decay\\ SNR (dB)\end{tabular} &
  \begin{tabular}[c]{@{}l@{}}1\\ 64\\ 16\\ 4\\ 10\end{tabular} \\ \hline
Classifier &
  Types of Classifiers &
  Mobilenetv2, TextCNN \\ \hline
SE Module &
  \begin{tabular}[c]{@{}l@{}}Num of Conv layers of the Encryption Module\\ Num of Conv layers of the Decryption Module\\ Num of Conv layers of the Reduction Module\end{tabular} &
  \begin{tabular}[c]{@{}l@{}}4\\ 4\\ 4\end{tabular} \\ \hline
PAJ Module &
  Num of linear layers of ATN &
  6 \\ \hline
SR Module &
  \begin{tabular}[c]{@{}l@{}}Num of ConvTranspose layers of SemGenerators\\ Num of Conv layers of SemDiscriminator\\ The dimension of the random vector\end{tabular} &
  \begin{tabular}[c]{@{}l@{}}3\\ 4\\ 100\end{tabular} \\ \hline
Training &
  \begin{tabular}[c]{@{}l@{}}Batch size\\ Learning rate\\ Optimizer\end{tabular} &
  \begin{tabular}[c]{@{}l@{}}256\\ 1.00E-04\\ Adam\end{tabular} \\ \hline
\end{tabular}
}

\end{table}

\subsection{Experimental Settings}
\par{We conduct experiments on a server with Ubuntu 18.04 operating system, equipped with 90GB RAM and an RTX3090 GPU card. The Pytorch and Python versions are 1.7.0 and 3.7, respectively. For drivers, the CUDA and cuDNN versions installed in the server are 11.0 and 8004, respectively. The parameters of neural networks are randomly initialized and then iteratively updated by the Adam optimizer. Table \ref{tab:param} shows the detailed parameters for models used in SemProtector.}
\begin{figure}[!t]
  \centering
  \includegraphics[width=72mm]{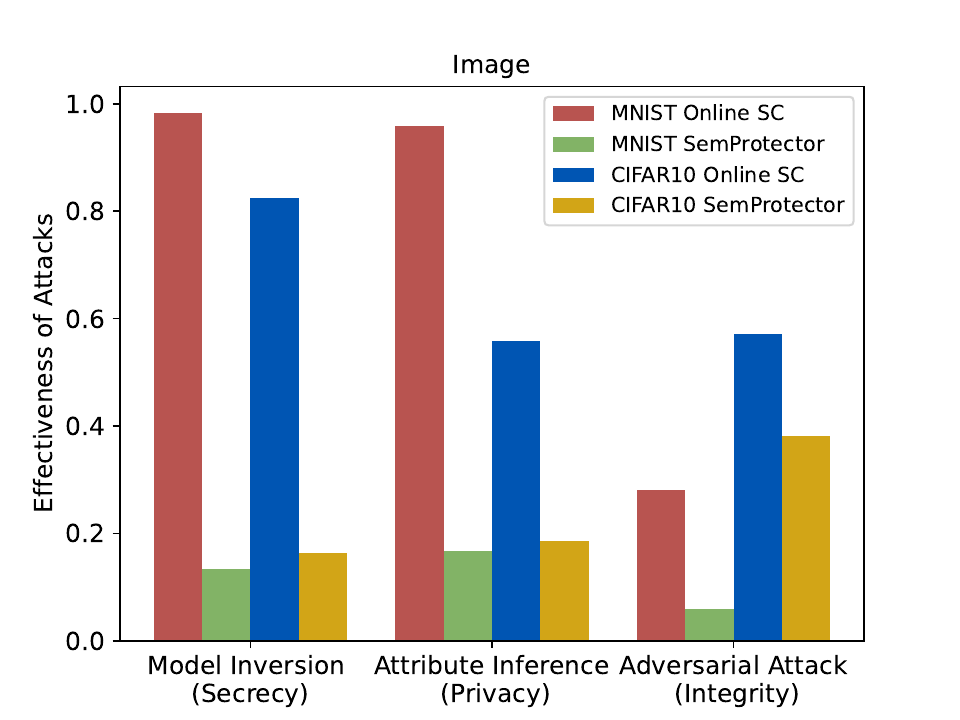}
  \captionof{figure}{Protection effect on image.}
  \label{fig:MNIST}
\end{figure}

\begin{figure}[!t]
  \centering
  \includegraphics[width=75mm]{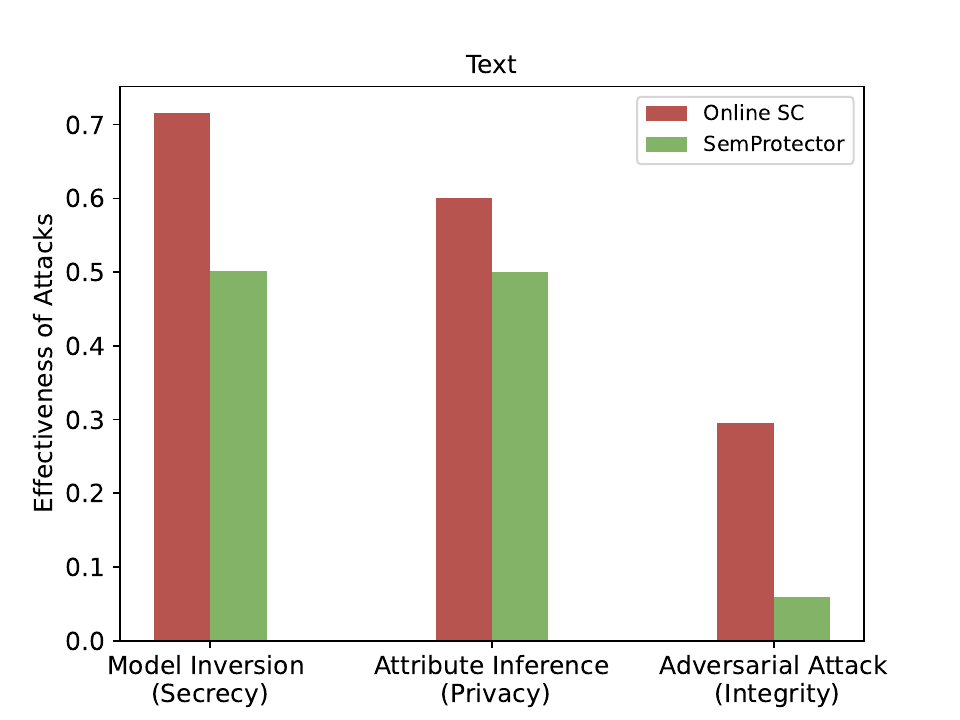}
  \captionof{figure}{Protection effect on text.}
  \label{fig:Text}
\end{figure}
\subsection{Main Results}

Fig.\ref{fig:MNIST} and Fig.\ref{fig:Text} demonstrate the effectiveness of our SemProtector on both image and text datasets, respectively. We assemble the three modules, i.e., SemEryp, SemPriv, and SemRect, to harden the online SC system against various attacks. The results show that the proposed SemProtector can significantly reduce the injuries under various destructive threats on both text and image modalities. Fig.\ref{fig:MNIST} shows that an eavesdropper can use the model inversion attacks to successfully infer the semantic data of an MNIST image, achieving as high as $98.3$\% accuracy based on its semantic symbols over the air, while the attacker can only obtain a $13.4$\% accuracy for the online SC system under the protection of our SemProtector. Fig. \ref{fig:Text} shows that our framework can  effectively protect the text-based SC system against various attacks, e.g., an adversarial attacker can only achieve a $6$\% successful rate equipped with our SemProtector. These results confirm the effectiveness of our framework in protecting semantic confidentiality. We also observe that the protections on MNIST are better than the ones on CIFAR10. One possible reason is that the semantics in the images of CIFAR10 are much more complex than the ones in MNIST.

We also quantify the side effects of our SemProtector to the image reconstruction for legitimate users. Table \ref{tab:Impact} reports the performance of our SemProtector under various settings. For example, ``Online SC + SemPriv'' indicates that the SemPriv module is plugged into our online SC system, and ``Online SC + SemEryp + SemPriv + SemRect'' represents that we assemble all three modules to harden the communication system. The results show that our SemProtector can lead to slight performance decreases in terms of SSIM and PSNR on the two datasets. It is reasonable that there will be a larger performance decrease with more protection modules involved. For customized protection requirements, one may need a trade-off between the security level of semantic  secrecy, privacy, integrity, and transmission quality.  

\begin{table}[htbp]
\caption{The impact of pluggable semantic protection modules on our SC system.\label{tab:Impact}}
\begin{center}
\resizebox{\linewidth}{!}{
\begin{tabular}{@{}ccccc@{}}
\toprule
\multicolumn{1}{l}{} & \multicolumn{2}{c}{\textbf{MNIST}} & \multicolumn{2}{c}{\textbf{CIFAR10}} \\ 
\multicolumn{1}{l}{} & PSNR & SSIM & PSNR & SSIM  \ \  \\
\midrule
Online SC                                & 27.88 & 0.99 & 25.49 & 0.87 \ \  \\
Online SC + SemEryp                      & 24.09 & 0.94 & 22.91 & 0.78 \ \  \\
Online SC + SemPriv                      & 27.75 & 0.97 & 25.04 & 0.86 \ \  \\
Online SC + SemRect                      & 26.69 & 0.93 & 23.19 & 0.77 \ \  \\
Online SC + SemPriv + SemRect            & 26.67 & 0.93 & 23.08 & 0.76 \ \  \\ 
Online SC + SemEryp + SemPriv            & 24.06 & 0.93 & 22.94 & 0.79 \ \  \\
Online SC + SemEryp + SemRect            & 23.87 & 0.92 & 22.90 & 0.76 \ \  \\
Online SC + SemEryp + SemPriv + SemRect  & 23.77 & 0.92 & 22.85 & 0.76 \ \  \\
\bottomrule
\end{tabular}
}
\end{center}

\end{table}

\section{Future Directions}
Our SemProtector addresses some fundamental challenges of securing semantics for online SC systems over open wireless channels. We further discuss two important research directions for the security of semantic communications.

\subsection{Certified robustness of semantic communications}
Although our SemProtector secures online SC with a unified framework and significantly hardens the system against various attacks, we still lack a theoretical analysis under what conditions the system keeps stable. A further step we can take is certified robustness, which provides the lower bound of robust accuracy against attacks under certain conditions. Robustness verification holds the promise of building reliable the end-to-end SC systems with a theoretical guarantee. Compared to traditional certified robustness, we need to additionally consider the physical layer adversarial attacks in fluctuated wireless channels, as well as the adversarial perturbations to the source data to derive a tight bound. These studies are still at an early stage, and developing certifiably robust approaches for an online SC system is still an open research problem.

\subsection{Security of distributed semantic communications}
Massive machine-type communications (mMTC) proposed in 5G is one of the key enablers for future cellular-based factory automation, using smart sensors and actuators to enhance manufacturing and industrial processes. Distributed semantic communications facilitate the above industrial IoT applications with a highly reliable and minimal latency communication paradigm. However, we will face challenging security issues due to the massive connected devices that can sense, communicate and store information about themselves. These devices are deployed in heterogeneous wireless networks and need to make intelligent decisions, and hence they are more fragile to various attacks. Securing distributed semantic communications in industrial IoT remains an open research problem.

\section{Conclusion}
\par{This paper introduces SemProtector, a unified framework that aims to secure an online SC system with three hot-pluggable semantic protection modules. These protection modules are able to encrypt semantics, mitigate privacy risks from wireless channels and calibrate distorted semantics at the destination. Our framework can also enable an existing online SC system to dynamically assemble the three modules to meet customized semantic protection requirements, greatly facilitating the practical deployment in real-world SC systems. Although SemProtector is tested on an image-based semantic communication system, we believe that our framework can also be further extended to video-based SC systems.}
\bibliographystyle{IEEEtran}
\bibliography{papercite.bib}
\section*{Biographies}
\vspace{-10 mm}
\begin{IEEEbiographynophoto}{Xinghan Liu}
(liuxinghan\_2022@bupt.edu.cn)  is currently pursuing a master's degree in National Engineering Research Center for Mobile Network Technologies, Beijing University of Posts and Telecommunications, Beijing 100876, China. His research interests include semantic communication security and privacy preservation.
\end{IEEEbiographynophoto}
\vspace{-10 mm}
\begin{IEEEbiographynophoto}{Guoshun Nan}
(nanguo2021@bupt.edu.cn) is a professor of National Engineering Research Center for Mobile Network Technologies, Beijing University of Posts and Telecommunications, Beijing 100876, China. He has broad interest in semantic communications, machine learning and wireless communications. He has published over 30 papers in top-tier conferences and journals, including ACL, CVPR, KDD, IEEE Network, and IEEE JSAC.
\end{IEEEbiographynophoto}
\vspace{-10 mm}
\begin{IEEEbiographynophoto}{Qimei Cui}
(cuiqimei@bupt.edu.cn) received the B.E. and M.S. degrees in electronic engineering from Hunan University, Changsha, China, in 2000 and 2003, respectively, and the Ph.D. degree in information and communications engineering from the Beijing University of Posts and Telecommunications (BUPT), Beijing, China, in 2006. She has been a Full Professor with the School of Information and Communication Engineering, BUPT, since 2014. Her research interests include B5G/6G wireless communications, edge computing and IoT.
\end{IEEEbiographynophoto}
\vspace{-10 mm}
\begin{IEEEbiographynophoto}{Zeju Li}
(lizeju@bupt.edu.cn) is an undergraduate student at Beijing University of Posts and Telecommunications (BUPT), Beijing, China. His research interests include AI safety, NLP and semantic communications.
\end{IEEEbiographynophoto}
\vspace{-10 mm}
\begin{IEEEbiographynophoto}{Peiyuan Liu}
(pyuanL@bupt.edu.cn) is currently pursuing a master's degree in Beijing University of Posts and Telecommunications (BUPT). Her research interests include federated learning security and semantic communications.
\end{IEEEbiographynophoto}
\vspace{-10 mm}
\begin{IEEEbiographynophoto}{Zebin Xing}
(xzebin@bupt.edu.cn) is an undergraduate student at Beijing University of Posts and Telecommunications (BUPT), Beijing, China. His research interests include Federal Learning and Differential privacy
\end{IEEEbiographynophoto}
\vspace{-10 mm}
\begin{IEEEbiographynophoto}{Hanqing Mu}
(muhanqing@emails.bjut.edu.cn) is currently pursuing a master's degree in Beijing University of Posts and Telecommunications (BUPT). His research interests include AI safety and privacy preservation.
\end{IEEEbiographynophoto}
\vspace{-10 mm}
\begin{IEEEbiographynophoto}{Xiaofeng Tao}
(taoxf@bupt.edu.cn) received the B.S. degree in electrical engineering from Xi’an Jiaotong University, Xi’an, China, in 1993, and the M.S. and Ph.D. degrees in telecommunication engineering from Beijing University of Posts and Telecommunications(BUPT), Beijing, China, in 1999 and 2002, respectively. He is a Professor in BUPT, a Fellow of the Institution of Engineering and Technology, and Chair of the IEEE ComSoc Beijing Chapter. He has authored or co-authored over 200 papers and three books in wireless communication areas. He focuses on 5G/B5G research.
\end{IEEEbiographynophoto}
\vspace{-10 mm}
\begin{IEEEbiographynophoto}{Tony Q.S. Quek}
(tonyquek@sutd.edu.sg) received the B.E. and M.E. degrees in Electrical and Electronics Engineering from Tokyo Institute of Technology, respectively. At MIT, he earned the Ph.D. in Electrical Engineering and Computer Science. Currently, he is the Cheng Tsang Man Chair Professor and Full Professor with Singapore University of Technology and Design (SUTD). He is currently serving as an Editor for the IEEE Transactions on Wireless Communications as well as an elected member of the IEEE Signal Processing Society SPCOM Technical Committee. He was an Executive Editorial Committee Member of the IEEE Transactions on Wireless Communications, an Editor of the IEEE Transactions on Communications. He is a Fellow of IEEE and a Fellow of the Academy of Engineering Singapore.
\end{IEEEbiographynophoto}
\vfill
\end{document}